# Semantifying Twitter: the influenceTracker ontology

Gerasimos Razis, Ioannis Anagnostopoulos,
Computer Science and Biomedical Informatics Dpt.
University of Thessaly
Lamia, Greece
{razis, janag}@dib.uth.gr

*Abstract*: **In this paper, we propose an ontology schema towards semantification provision of Twitter social analytics. The ontology is deployed over a publicly available service that measures how influential a Twitter account is, by combining its social activity and interaction over Twittersphere. Apart from influential quantity and quality measures, the service provides a SPARQL endpoint where users can perform advance semantic queries through the RDFized Twitter entities (mentions, replies, hashtags, photos, URLs) over the semantic graph.**

## 1. INTRODUCTION

Microblogging is a form of Online Social Network (OSN) which attracts millions of users on daily basis. Twitter is one of these microblog services. Their users vary from citizens to political persons and from news agencies to huge multinational companies. Obviously, some users are more influential than others. A service is required for quantifying and measuring the value of that influence. We have created InfluenceTracker[1], a publicly available website[1] where anyone can rate and compare the recent activity of any Twitter account.

The aim of this paper is twofold. Firstly, we propose an improvement over a previous work of us, which is used for calculating the importance and influence of a Twitter account. This improvement incorporates a quality measurement that reflects other users' "opinion" and preference over the examined tweets. Secondly, we aim to describe a proposed ontology and the related semantic mechanisms/technologies which will allow us to semantify the disseminated information and the respective Twitter entities (mentions, replies, hashtags, photos, and URLs) as Linked Data.

The remainder of this paper is organized as follows. In the next section, we provide an overview over the related work on discovering influential users and on generating RDF graphs from Twitter data. Then, in Section III we describe the introduced improvement for the calculation of the importance of Twitter accounts. In Section IV we analytically present the implemented on-line service and the ontology employed for the transformation of the Twitter accounts and their entities (mentions, replies, hashtags, photos, URLs) into an RDF graph. Finally, Section V provides the conclusions of our work by summarizing the derived outcomes, while providing considerations on our future directions.

## 2. RELATED WORK

The calculation of the impact a user has on social networks, as well as the discovery of influencers in them is not a new topic. It covers a wide range of sciences, ranging from sociology to viral marketing and from oral interactions to Online Social Networks (OSNs). In the related literature, the term "influence" has several meanings and it is differently considered most of the times.

The authors in [1] utilize a large number of tweets containing at least one URL, their authors and their followers. Their aim is to calculate how influential or passive the Twitter users are. The produced influence metric depends on the "Follower-Following" relations of the users, as well as their retweeting behavior. The authors state that the number of followers a user has, is a relatively weak predictor of the maximum number of views a URL can achieve. As our work has shown [2], through the retweet functionality information can be diffused to audience not targeted.

The work described in [3] proposes a methodology where for each Twitter user three different types of influence are introduced. These types are "Indegree" (number of followers), "Retweet" (number of user generated tweets that have been reweeted) and "Mention" (number of times the user is mentioned in other users' tweets). A necessary condition for the computation of these influence types is the creation of at least ten tweets per user. The authors claim that "Retweet" and "Mention" influence correlate well with each other, while the "Indegree" does not. Therefore, they come up with the conclusion that users with high "Indegree" influence are not necessarily influential.

---

[1] http://influencetracker.com/

A topic-oriented study on the calculation of influence in OSNs is presented in [4]. The authors propose an algorithm which takes into consideration both the topical similarity between users and their link structure. It is claimed that due to homophily, which is the tendency of individuals to associate and bond with others having similar interests, most of the "Follower-Following" relations appear. This work also suggests that the active users are not necessarily influential.

Another approach which defines influence in terms of copying what the directly related do is presented in [5]. In this work, the authors propose an "influenceability" score, which represents how easily a user is influenced by others or by external events. It is built on the hypothesis that a very active user performs actions without getting influenced by anyone. The users of such a type are considered as responsible for the overall information dissemination in the network.

The study in [6] shows that retweeting can be also characterized as a conversational infrastructure. According to the authors, a conversation "exists" either during a retweet where some new information is added to the initial message, or when a single tweet is retweeted multiple times. The latter is interpreted by the authors as an action to invite new users into the conversation.

The work presented in [7] introduces a semantic data aggregator, which combines a collection of compact formats for structured microblog content with Semantic Web vocabularies. Its main purpose is to provide user-driven Linked Data. The main focus of this work is on microblog posts and specifically on their creators, their content and their associated metadata.

Another framework which utilizes semantic technologies, common vocabularies and Linked Data in order to extract and mine microblogging data regarding scientific events from Twitter is proposed in [8]. The authors attempt to identify persons and organization related to them based on entities of time, place and topic.

All the related studies have shown that the most active users or those with the most followers are not necessarily the most influential. This fact has also been spotted by our work. Our Influence Metric depends on a set of factors (Section III), while the account activity is only one of them.

Contrary to the aforementioned studies, for the calculation of our Influence Metric we neither set a lower threshold on the number of the user-generated tweets, nor we only utilize a specific subset of tweets that fulfill certain criteria (e.g. those containing URL etc.). Thus, all the Twitter accounts can be used as seed for the calculation of our Influence Metric.

## 3. DEFINING INFLUENTIAL METRICS

Twitter accounts are the building blocks of that Social Network. If depicted in a graph, accounts are represented by nodes. Edges that connect these nodes are the relations of "Follower-Following". Obviously, some accounts are more influential than others. The methodology of calculating the importance and influence of a Twitter account has been presented in [2]. In this paper, we present an improved version of this metric, by incorporating a new factor.

As mentioned above, the influence measurement should not depend merely on the number of "Followers", even if that number is big enough. In case that the number of "Following" is larger, then the user could be characterized as a "passive" one. That type of users are regarded as those who are keener on viewing or being informed through tweets rather than composing new ones. Therefore, a suitable factor is the ratio of "Followers to Following" (*FtF* ratio). However, this ratio is also not sufficient. Another important factor is the tweets creation rate (*TCR*). For example, let us see the case where two accounts have nearly the same *FtF* ratio. Obviously, the account with the higher *TCR* has more impact on the Network. In our methodology, in order to calculate that rate, we process the latest 100 tweets of the account according to the Twitter API. That helps us to keep dynamic the values of *TCR* (and consequently the Influence Metric), as it depends on the most recent activity of the accounts in Twitter. In order to improve the precision of this metric, the timeframe of its calculation is no longer measured in days, but rather in hours.

Each tweet is associated with several other kinds of information presented in influencetracker.com. Two of them are the "Retweets" and "Favorites" counts, which represent how many times a Tweet has been retweeted as well as marked as favorite by other users respectively. In our methodology, we utilize these counts in order to calculate the h-index of the "Retweets" and "Favorites", over the last 100 tweets of an examined account. The aim of these measurements is to provide a quality overview of the tweets of a Twitter account in terms of likeability and impact in the Twittersphere. These indexes are based on the established h-index [9] measurement and are named *"ReTweet h-index - Last 100*

*Tweets"* and *"Favorite h-index - Last 100 Tweets"*. The most important factor regarding them is that they reflect other users' assessment of the content of the tweets.

Consequently, a Twitter account has *"ReTweet h-index - Last 100 Tweets"* equal to *h*, if *h* over the last *Nt* tweets have at least *h* retweets each, and the remaining *(Nt - h)* of these tweets have no more than *h* retweets each (max. *Nt*=100). This can be interpreted as follows: at least *h* tweets have been retweeted at least *h* times. Thus, we consider that this retweeting action results in the generation of at least *h\*h* new tweets, which have to be attributed to the account that initially posted them.

TABLE I. CALCULATING THE "ADJUSTED TWEETS"

| RT *h*-index | *h\*h* | Transformed as | Calculation Process | Adjusted Tweets |
|---|---|---|---|---|
| 0,3 | - | 0,3 * 10^0 | 0,3/10, 10^0 | 0,03 |
| 2 | 4 | 4 * 10^0 | 4/10, 10^0 | 0,4 |
| 6 | 36 | 36 * 10^0 | 36/10, 10^0 | 03,6 |
| 15 | 225 | 22,5 * 10^1 | 22/10, 10^1 | 12,2 |
| 45 | 2.025 | 20,25 * 10^2 | 20/10, 10^2 | 22 |
| 80 | 6.400 | 64 * 10^2 | 64/10, 10^2 | 26,4 |
| 100 | 10.000 | 10 * 10^3 | 10/10, 10^3 | 31 |

However, prior to incorporating this amount of new tweets into the equation of the Influence Metric, we employ a calculation mechanism for avoiding outliers. Moreover, we introduce a value called "Adjusted Tweets" which is defined in Equation 1.

$$\text{Adjusted Tweets} = a \times 10^b, \text{ where } 0 < a < 100 \text{ and } a \in \mathcal{R} \quad (1)$$

"Adjusted Tweets" are actually a form of expressing the *h\*h* value. Where applicable, "a" is a two-digit number. Then, that number is divided by 10. The resulting quotient is combined with the Order of Magnitude of the *h\*h*, which is represented by "b", thus forming the "Adjusted Tweets Number" according to Equation 1. Some characteristic examples are provided in Table I.

As already mentioned, the tweets generated from the retweeting process have to be attributed to the account that initially posted them. Therefore, the value of the "Adjusted Tweets" is added to the 100 tweets retrieved from the account, as defined in Equation 2. The *FtF* ratio is placed inside a base-10 log for avoiding outlier values. Moreover, this ratio is added by 1, so as to avoid the metric being equal to 0 in case where the values "Followers" and "Following" are equal.

$$\text{Influence Metric} = \frac{\text{tweets}_k + \text{AdjustedTweets}_k}{\text{Hours}_{\text{since } k_{\text{th}}\text{tweet}}} *$$

$$* \text{OOM}(\text{Followers}) * \log_{10}\left(\frac{\text{Followers}}{\text{Following}} + 1\right),$$

where OOM: Order Of Magnitude (2)

In an effort to compare our metric with other approaches we used Followerwonk[2], which is a well-known service that provides the "Social Authority" value (a similar measure of how influential an account is). Thus, in order to directly evaluate our metric, we randomly selected nearly 11.500 Twitter accounts and then we compared the ranking positions as well as the average ranking differences in all positions levels for both services (InfluenceTracker and Followerwonk). In Figure 1 (in the Appendix), the horizontal-axis value of each point corresponds to the InfluenceTracker ranking position the tested account receives, while the vertical-axis value corresponds to the respective ranking the account receives from Followerwonk. We observed that the average position ranking difference is 1476 (the black linear trendline is defined by the function *y=0.783\*x+1476)*. Finally, in comparison to the ideal curve (red line that is defined by *y=x*) we noticed that outlying values are equally distributed between the higher and the lower ranking positions assigned by our service.

---

[2] http://followerwonk.com/

## 4. OUR PROPOSED SERVICE

In this section, we present the architecture and infrastructure of influencetracker.com service. In addition, we describe the ontology used for transforming the Twitter accounts and the entities of the examined tweets (mentions, replies, hashtags, photos, URLs) into an RDF graph.

*A. Infrastructure*

The architecture of the influencetracker.com service and the relevant data flows are presented in Figure 3 (in the Appendix). The service combines the use of a relational database joint with an RDF triple store. Thus, data and related information displayed at the web pages combine both technologies. The relational database is a MySQL Server and the RDF triple store is contained in an Open Link Virtuoso (OLV) Server. There are two use case scenarios of the service.

The first use case involves the updating process of the RDF graph. A service, which is implemented using Python libraries, is executed every four days. The process is split into four phases. During the first phase, a request is sent to the Twitter API for each account found in the database. The response contains the data in JSON format. In the second phase the necessary data are parsed and the metrics are calculated. The third phase involves the semantification of the gathered data with concepts (resources and property URIs) derived from our ontology (see Figure 2) and RDF graph updates. This process is performed by using the RDFLib framework. During the last phase, the triples are stored in the OLV environment, while the user can use a SPARQL endpoint[3,4] for semantic search. Related processes are presented in Figure 1.

The second use case is a subset of the aforementioned one. It takes place when a Twitter account is searched through the provided web interface[5]. Another service, also implemented in Python, performs a request to the Twitter API for the investigated account. A response is returned in JSON format. In case of a valid account, the necessary data are parsed, the metrics are calculated and stored in the relational database. In this use case, no data are stored at the RDF graph. This is because we wanted to maximize the responsiveness of our service, minimizing in parallel the execution time. Finally, in case where a new account is inserted into our system, the necessary data will be stored at the RDF graph during the next update process.

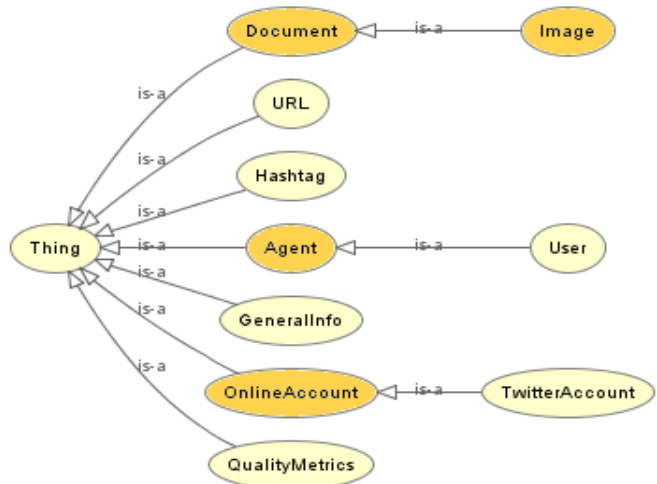

Fig. 2. The hierarchy of the classes of the "InfluenceTracker" ontology

*B. The "InfluenceTracker" Ontology*

Another aim of this work is to model the Twitter accounts and their related information in tweets (e.g. mentions, replies, URLs, hashtags, photographs). This ontology utilizes properties from the FOAF ontology [10]. FOAF (Friend-of-a-Friend) is an ontology for describing persons, their activities as well as their relations to other people and objects, while it can be generalized as to describe all type of entities, called agents, who are responsible for specific actions [10]. In our context the agents are the Twitter users, who are responsible for specific actions, such as owing Twitter accounts, posting tweets, etc. Figure 2 displays the classes and their hierarchical relationships. During the representation of the entities, two specific prefixes are used, namely "foaf" and "it". They correspond respectively to the namespace of the FOAF and of our proposed ontology. The ontology is built on three basic building blocks: the classes, the object properties and the datatype ones.

---

[3] http://www.influencetracker.com/endpoint
[4] http://www.influencetracker.com:8890/sparql
[5] http://www.influencetracker.com/searchAccount

  1) Classes

The classes are used to represent conceptual entities. Those defined in the influenceTracker ontology are the following:
- foaf:Agent: A general class which describes agents who are responsible for several actions [10].
- it:User: It is a subclass of the foaf:Agent and describes the agents that own a Twitter account. These may be physical persons, organizations or events.
- foaf:OnlineAccount: It represents the provision of some form of online service, by some party (indicated indirectly via the foaf:accountServiceHomepage object property) to some foaf:Agent [10].
- it:TwitterAccount: The class is a subclass of the foaf:OnlineAccount and represents the actual Twitter accounts.
- it:GeneralInfo: The class contains the Twitter related details of an account characterized by influencetracker.com as "General Information". These are the total number of tweets, the *TCR*, the retweet ratio, and the number of followers and following.
- it:QualityMetrics: The class contains the metrics of a Twitter account characterized by influencetracker.com as "Quality Metrics". These are the *"ReTweet and Favorite h-index - Last 100 Tweets"*, the estimated *"ReTweet and Favorite h-index"*, the reply ratio and the value of our influence metric.
- foaf:Document: The class represents those things which are, broadly conceived, documents. There is no distinction between physical and electronic ones [10].
- foaf:Image: The class corresponds to those documents which are images. It is a subclass of the foaf:Document, since all images are documents. Digital images are instances of this class [10].
- it:Hashtag: The class describes the entities which are hashtags (words starting with "#").
- it:URL: The class describes the entities which are URLs.

  2) Object Properties

The object properties are those for which their value is an individual. Those defined in the influenceTracker ontology along with their concept restrictions are the following:
- foaf:account: The property is used to relate a foaf:Agent to a foaf:OnlineAccount for which they are the sole account holder [10].
- foaf:accountServiceHomepage: The property indicates a relationship between a foaf:OnlineAccount and the homepage of the supporting service provider [10].
- it:hasGeneralInfo: The property relates an it:User to an it:GeneralInfo which contains the Twitter related information of the owned account, characterized by influencetracker.com as "General Information".
- it:hasMentioned: The property relates an it:User to an it:User that has been mentioned in the first user's tweets.
- it:hasQualityMetrics: The property relates an it:User to an it:QualityMetrics which contains the metrics of the owned account, characterized by influencetracker.com as "Quality Metrics".
- it:hasRepliedTo: The property relates an it:User to an it:User that has received a tweet as a reply from the first user.
- it:includedHashtag: The property relates an it:User to an it:Hashtag that has been included in the user's tweets.
- it:includedImage: The property relates an it:User to an it:Image that has been included in the user's tweets.
- it:includedUrl: The property relates an it:User to an it:URL that has been included in the user's tweets.

These properties have been defined in such a way so as to be easily extensible to cover concepts from other OSNs as well. The Twitter accounts can be replaced by those of Facebook and the tweets by the statuses. The actions of "Share" and "Like" found in Facebook are the equivalent of "Retweet" and "Favorite" of Twitter. The concepts of hashtags, mentions, replies, images and URLs are the same in these OSNs.

  3) Datatype Properties

The datatype properties are those for which their value is a data literal. Those defined in the influenceTracker ontology along with their concept restrictions are the following:

- foaf:accountName: The property provides a textual representation of the account name (unique ID) associated with that account [10].
- it:description: The property provides the description of an account, as set by its owner.
- it:displayName: The property provides the name displayed at the web page of an account, as set by its owner.
- it:followers: The property provides the number of the followers of an account.
- it:following: The property provides the number of the accounts that an account follows.
- it:hIndexFav: The property provides the value of the *"Favorite h-index - Last 100 Tweets"* metric of an account.
- it:hIndexFavDaily: The property provides the estimated daily value of the *"Favorite h-index"* metric during the lifespan of an account.
- it:hIndexRt: The property provides the value of the *"ReTweet h-index - Last 100 Tweets"* metric of an account.
- it:hIndexRtDaily: The property provides the estimated daily value of the *"ReTweet h-index"* metric during the lifespan of an account.
- it:imageUrl: The property provides the URL that leads to an image which was included in a tweet.
- it:influenceMetric: The property provides the value of the Influence Metric measurement. Its aim is to describe both the importance and impact of an account in a social network.
- it:profileLocked: The property indicates whether the profile of an account is publicly visible or not.
- it:replyRatio: The property provides the ratio of the user's latest tweets which are used as replies to other users' tweets.
- it:retrievedOn: The property provides the date that the information regarding an account was lastly updated.
- it:rtPercent: The property provides the percentage of the latest user's tweets that are retweets from other accounts.
- it:tweets: The property provides the number of the total tweets posted by an account.
- it:tweetsPerDay: The property provides the number of the average tweets posted per day by an account.
- it:url: The property provides the URL that leads to a web site which was included in a tweet.

*C. SPARQL Queries*

As already mentioned, a public endpoint allows the search of the collected semantic data. The following SPARQL query returns the Twitter accounts (along with their full details), which were mentioned and at the same time have received a reply by a specific account {accountName}. Another dependency is the existence of the details of these accounts in the RDF graph.

```
PREFIX it: <http://www.influencetracker.com/ontology#>
PREFIX foaf: <http://xmlns.com/foaf/0.1/>
SELECT ?twitterAccount ?accountName ?influenceMetric ?h_index_RT ?h_index_Fav ?tweetsNum ?following
?followers ?TPD ?rtPerc ?h_index_RT_Daily ?h_index_Fav_Daily ?reply_ratio
WHERE {
    <http://www.influencetracker.com/resource/User/{accountName}> it:hasMentioned ?mentioned .
    ?mentioned foaf:account ?twitterAccount ;
            it:hasQualityMetrics ?quality ;
            it:hasGeneralInfo ?general .
    ?twitterAccount foaf:accountName ?accountName .
    ?quality it:influenceMetric ?influenceMetric ;
          it:hIndexRt ?h_index_RT ;
          it:hIndexFav ?h_index_Fav ;
          it:hIndexRtDaily ?h_index_RT_Daily ;
          it:hIndexFavDaily ?h_index_Fav_Daily ;
          it:replyRatio ?reply_ratio .
    ?general it:tweets ?tweetsNum ;
          it:following ?following ;
          it:followers ?followers ;
          it:tweetsPerDay ?TPD ;
          it:rtPercent ?rtPerc .
    FILTER EXISTS {<http://www.influencetracker.com/resource/User/{accountName}> it:hasRepliedTo ?mentioned}
```

}
*ORDER BY DESC (?influenceMetric)*

The URIs which are returned by the queries are dereferenceable ones, consequently the resources that they identify are represented by documents, which in our case are in HTML format. Specifically, these URIs are constructed using the Slash format. An example of such URI is: "www.influencetracker.com/resource/User/youtube". It repre-sents the document where the resource "youtube", an instance of the "it:User" class, is described.

## 5. COCNCLUSIONS AND FUTURE WORK

The content of this paper has spanned over two areas. Firstly, an improvement on the existing Influence Metric was presented, by incorporating a quality metric. It is based on the established *h*-index measurement and it reflects other users' actions and preferences over the content of the created tweets. Secondly, an ontology which was used as the base for the semantification of the Twitter accounts and their entities (mentions, replies, hashtags, photos, and URLs) was described. Information is inserted into an RDF graph, which is publicly available for querying through a provided SPARQL endpoint. To the best of our knowledge, there is currently no active service for providing such kind of data.

In the future, we plan to expand our ontology by incorporating others (as we did in this work with FOAF) in order to cover more information, such as the "followers-following" relationships introduced by Twitter. Such an expansion can be also used to capture similar kind of information from other OSNs. Twitter accounts can be replaced by those of Facebook and the tweets by the statuses. The actions of "Share" and "Like" found in Facebook are the equivalent of "Reteweet" and "Favorite" of Twitter.

APPENDIX

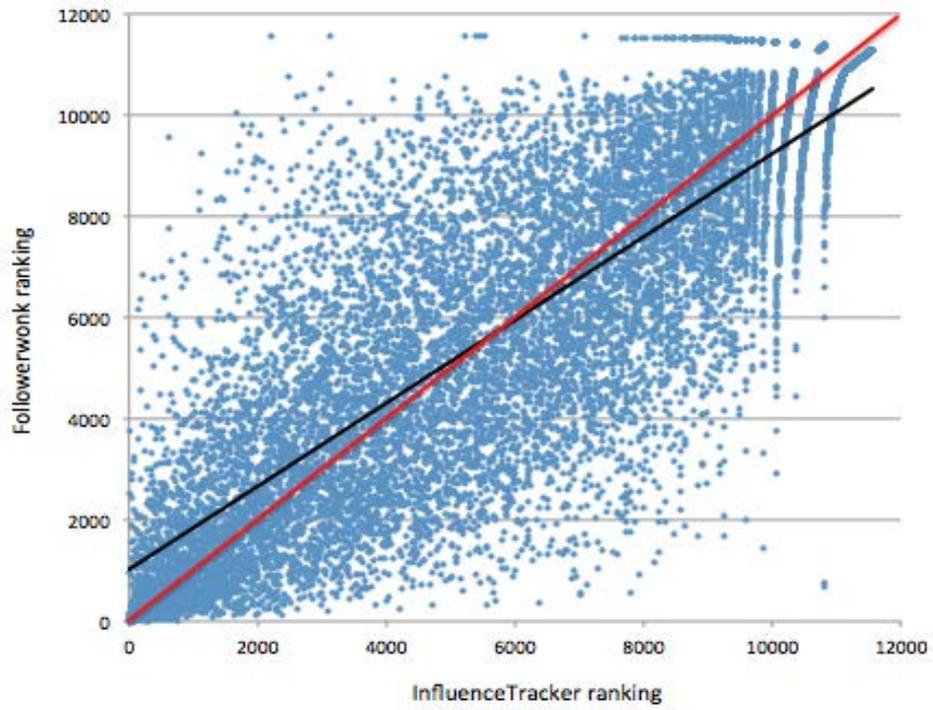

Fig. 1. Evaluation of our metric in comparison to Followerwonk service

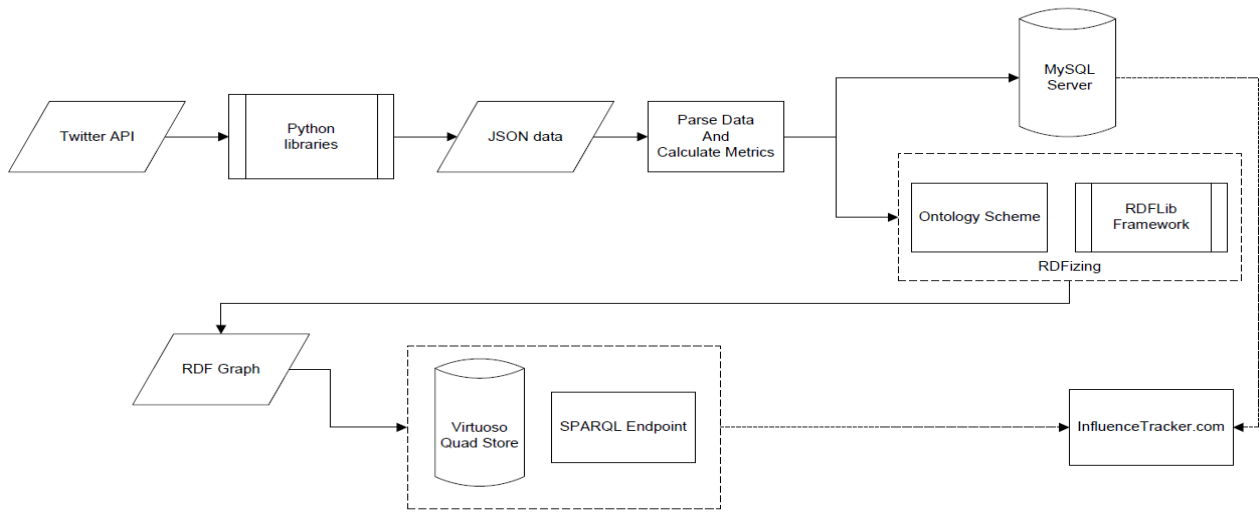

Fig. 3. The phases during the updating process of the RDF graph